\documentclass{iau}
\usepackage{graphicx}

\title[Bubbles S51, N68, and N131]{Triggered Star Formation \\from Bubbles S51, N68, and N131}

\author[C. P. Zhang \& J. J. Wang]   
{Chuan-Peng Zhang
 \and Jun-Jie Wang}

\affiliation{National Astronomical Observatories, Chinese Academy of
Sciences, \\ 100012 Beijing, PR China; email: {\tt
zcp0507@gmail.com}}

\pubyear{2012}
\volume{292}  
\pagerange{119--126}
 \jname{Molecular Gas, Dust, and Star Formation
in Galaxies} \editors{Tony Wong \& Juergen Ott Editor, eds.}
\begin{document}

\maketitle

\begin{abstract}
We investigated the environment of the infrared dust bubbles S51
\cite[(Zhang \& Wang 2012a)]{s51}, N68 \cite[(Zhang \& Wang
2012b)]{n68}, and N131 \cite[(Zhang et al. 2012)]{n131} from
catalogue of \cite[Churchwell et al. (2006)]{churchwll2006}, and
searched for evidence of triggered star formation.
\end{abstract}

\firstsection 

\begin{figure}[thb]
 \vspace*{0.5 cm}
\begin{center}
 \includegraphics[height=1.47in]{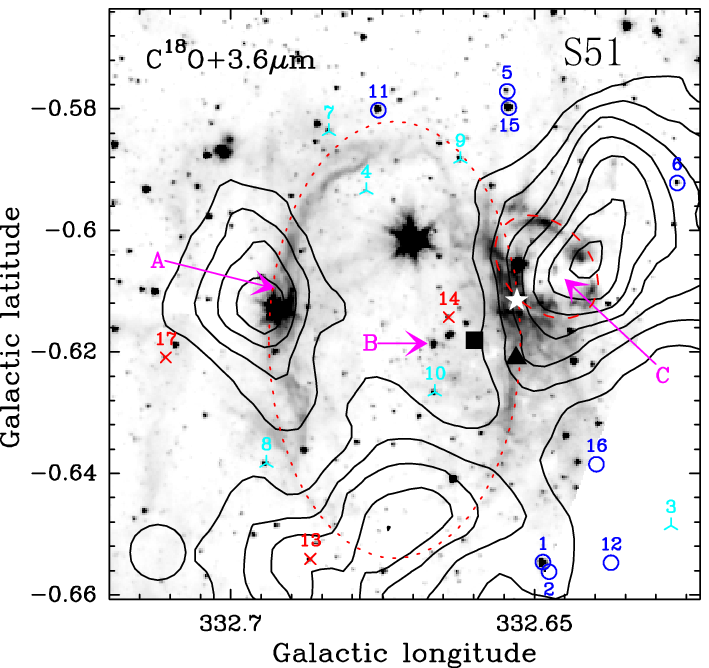}
  \includegraphics[height=1.47in]{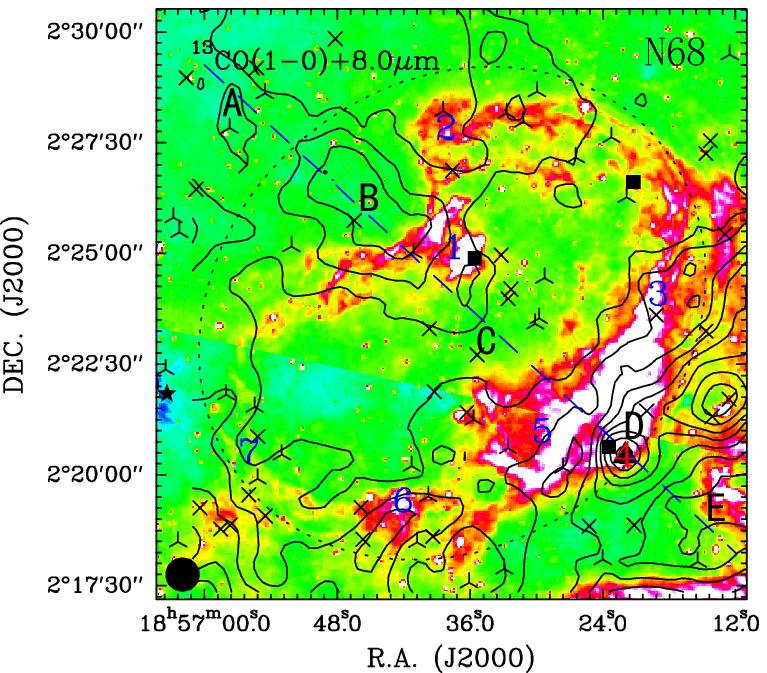}
  \includegraphics[height=1.48in]{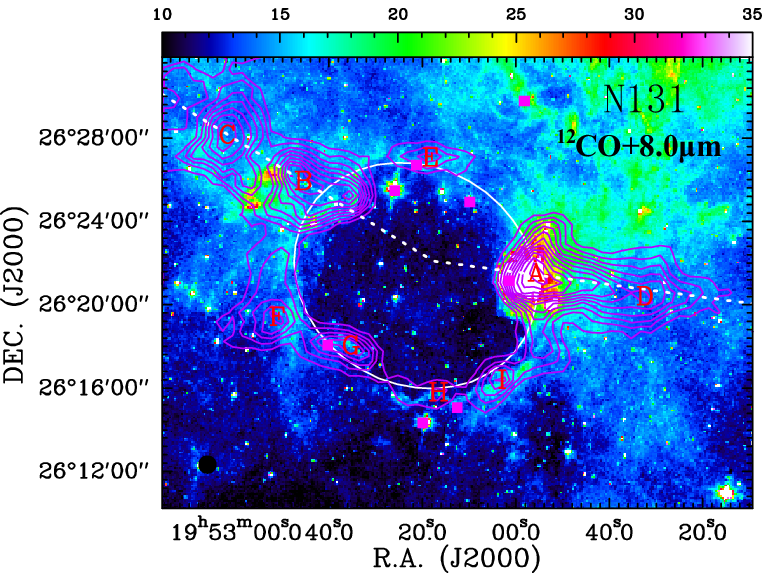}
\caption{Integrated velocity contours of the CO emission and young
stellar object (YSO) distribution superimposed on the GLIMPSE
infrared emission of three bubbles. Detailed information about these
three bubbles can be found in \cite[Zhang \& Wang 2012a]{s51},
\cite[Zhang \& Wang 2012b]{n68}, and \cite[Zhang et al.
2012]{n131}.}
   \label{fig1}
\end{center}
\end{figure}

The ringlike shell of the CO molecular gas is greatly correlative
with the infrared dust emission for the bubbles S51, N68, and N131.
The velocity distribution of Mopra $^{13}$CO and C$^{18}$O data
\cite[(Lo et al. 2009)]{lo2009} shows that the bubble S51 may be
made up of a front-side cloud and a ringlike shell. The
position-velocity diagram of GRS $^{13}$CO data shows that the
bubble N68 may be expanding outward at a speed of about 5 km
s$^{-1}$ in the line of sight. Integrated velocity contours and
velocity distribution of Delingha $^{12}$CO and $^{13}$CO data show
that there are two giant and collimated wings/flows of CO emission
appearing at the ringlike shell of the bubble N131. Morphologically,
it is very obvious that the ringlike shell and the two wings/flows
of the bubble N131 were triggered by the exciting star(s) inside
this bubble. In addition, several YSOs are located around the
ringlike shell of these bubbles, so YSOs may represent a second
generation of stars triggered by the bubble's expanding outward.

\end{document}